\documentclass[aps,prb,twocolumn,10pt,floatfix,eqsecnum,longbibliography,nofootinbib,showkeys,reprint,nobibnotes]{revtex4-2}

\usepackage{amsmath}
\usepackage{amssymb}
\usepackage{amsfonts}
\usepackage{tablefootnote}
\usepackage{graphicx}
\usepackage{color}
\usepackage[dvipsnames]{xcolor}
\usepackage{bbm}
\usepackage[breaklinks=true,colorlinks=true,linkcolor=blue,urlcolor=blue,citecolor=blue]{hyperref}
\usepackage{array}
\usepackage{booktabs}
\usepackage{mathptmx}
\usepackage{placeins}

\begin{document}
\title{Vortex ratchet effect in superconductor open nanotubes and nanopetals}

\author{Igor Bogush$^{1,2}$}
\email{igori.bogus@tu-braunschweig.de}
\author{Rodrigo H. de Bragan\c{c}a$^{3,4}$, Vladimir M. Fomin$^{3,5}$, Oleksandr V. Dobrovolskiy$^{1,2}$}
\affiliation{$^{1}$Cryogenic Quantum Electronics, Institute for Electrical Measurement Science and Fundamental Electrical Engineering, Technische Universit\"at Braunschweig, Hans-Sommer-Str. 66, 38106 Braunschweig, Germany}
\affiliation{$^{2}$Laboratory for Emerging Nanometrology, Technische Universit\"at Braunschweig, Langer Kamp 6A-B, 38106 Braunschweig,  Germany}
\affiliation{$^{3}$Institute for Emerging Electronic Technologies, Leibniz IFW Dresden,
 Helmholtzstra\ss e 20, 01069 Dresden, Germany}
\affiliation{$^{4}$Departamento de F\'{i}sica, Centro de C\^{e}ncias Exatas e da Natureza,
Universidade Federal de Pernambuco, Recife, PE, 50740-560, Brasil}
\affiliation{$^{5}$Faculty of Physics and Engineering, Moldova State University, strada A. Mateevici 60, MD-2009 Chi\c{s}in\u{a}u, Republic of Moldova}
 
\begin{abstract}
    Advancements in the fabrication of superconducting 3D nanostructures and the creation of artificial pinning sites pave the way to novel applications and enhancement of nanosensors, bolometers, and quantum interferometers. The dynamics of magnetic flux quanta (Abrikosov vortices) in 3D nanoarchitectures reveal a rich palette of phenomena unseen in planar counterparts. Here, we consider two types of superconductor 3D nanostructures -- open nanotubes and nanopetals -- carrying an azimuthal transport current in a homogeneous external magnetic field. The complex 3D geometry of the structures induces an inhomogeneity of the normal magnetic field and makes the vortices move along preferred paths. By introducing a series of asymmetric pinning sites along these paths, we demonstrate non-reciprocity in the flux transport, which, in the 3D nanostructures, is  stronger than in the planar membranes. The enhancement of the vortex ratchet effect manifests via a difference in the vortex depinning current under current reversal in a wider range of magnetic fields. The revealed effect is attributed to the inhomogeneous field-induced vortex channeling through the areas containing the asymmetric pinning sites. Our results demonstrate that the ratchet effect can persist up to higher magnetic fields via extending a superconducting film into the third dimension, without an increase in the number of asymmetric pinning sites.
\end{abstract}
\maketitle

\section{Introduction}

Superconductor 3D nanoarchitectures are a subject of vigorous theoretical and experimental studies because of fascinating superconducting properties and prospects for applications in nano- and optoelectronics, quantum optics, and information processing\,\cite{Thurmer10,Loe19acs,Mak21adm,Fom22apl,Cordoba24apl,zhakina2024vortex,Bog24prb,shani2020dna}. Notably, of particular interest are the effects arising from the extension of planar thin films into the third dimension and the influence of 3D geometry on phenomena previously studied in 2D planar films and constrictions.

Most application-relevant superconductors are known to be of type II. Under moderately high magnetic fields, they are penetrated by magnetic flux lines (Abrikosov vortices or fluxons). In certain applications, such as Abrikosov-vortex random access memory (AV-RAM) \cite{key-47}, vortices serve as information bits, while in others, the presence of moving vortices leads to unwanted dissipation \cite{key-55, key-15}. In order to prevent vortices from generating undesired voltage upon their motion, pinning sites are usually introduced \cite{Bra95rpp} to confine a vortex within a small region of the superconductor, of the order of the coherence length $\xi$. Then, in the presence of a transport current, a vortex experiences a driving Lorentz-type force $\mathbf{F}_{\text{L}}$ counterbalanced by a pinning force $\mathbf{F}_{\text{p}}$ created by the pinning sites. Besides, a vortex experiences a vortex-vortex repulsion and interacts with the energy barriers associated with the sample edges \cite{Bud22pra}. The length scale of this interaction is given by the magnetic field penetration depth $\lambda$.

\begin{figure*}
    \includegraphics[width=0.9\textwidth]{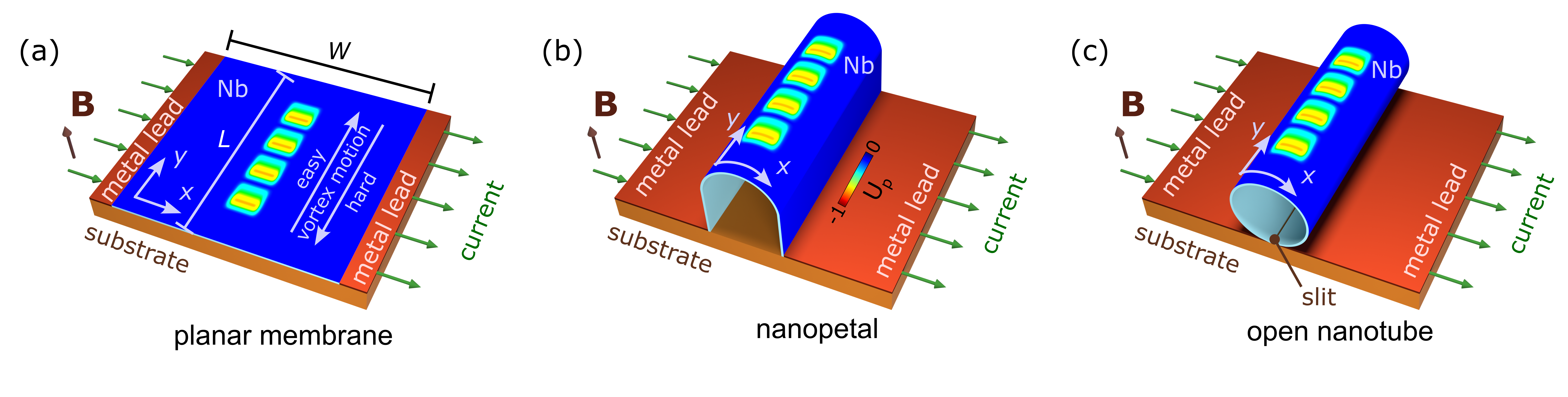}
    \caption{Geometries for a planar strip (a), a nanopetal (b), and an open nanotube (c), with the distribution of the pinning potential superimposed on the superconductor surface.}
    \label{fig:1}
\end{figure*}

The interaction between a vortex and a pinning site is modeled using a pinning potential. An asymmetry in the potential usually leads to a \textit{vortex ratchet effect} (VRE)\,\cite{Silhanek2010}, due to the different pinning forces acting on vortices moving in opposite directions. A VRE manifests as a rectified voltage under current reversal and an asymmetric current–voltage characteristics (CVC)\,\cite{Dob20pra,Silhanek2010,Plourde2009Pinning,Reichhardt2017ratchet}. Various pinning potential configurations were used to study VREs theoretically and experimentally, including random \cite{key-35} and ordered \cite{key-39, key-40} distributions, stacked arrowhead arrays \cite{key-17}, in-plane magnetized dipole arrays \cite{key-37}, and triangular antidot arrays \cite{key-38}.

Previously, by numerical modeling, we revealed  that the maximum normal magnetic field component, $\mathbf{B}_n$, at a convexity of open 3D superconductor nanotubes can act as a conveyor for vortices \cite{Bog24prb}. An interesting question is whether an inhomogeneous field-induced flux channeling through areas with asymmetric pinning sites can influence the VRE.

Here, we investigate the combined effects of 3D geometry and asymmetric pinning potential on vortex motion in open nanotubes and nanopetals, compared to flat strips. We demonstrate that the cooperative interaction between the inhomogeneous $\mathbf{B}_n$ field and the asymmetric pinning potential enhances the VRE in both the nanotube and nanopetal. We attribute this finding to the 3D geometry, which forces vortices to move through regions with asymmetric pinning sites, rather than between these regions, as observed in the reference planar structure. Numerical simulations are performed using the time-dependent Ginzburg-Landau (TDGL) equation, enabling us to obtain the current-voltage characteristics (CVCs) for the nanoarchitectures and to calculate the VRE efficiency by comparing depinning currents under current polarity reversal. Our theoretical predictions can be examined, for example, on superconducting 3D nanoarchitectures fabricated through strain-relaxation-driven self-rolling \cite{Thurmer08,Loe19acs}, with artificial pinning sites created using focused ion and electron beams \cite{Fer20mat,Vol24nac}.

\section{Model}
\subsection{Geometries under consideration}
The dynamics of vortices are studied for the three geometries shown in Fig.\,\ref{fig:1}: a flat strip (a), a nanopetal (b), and an open nanotube (c). The structures are placed in a homogeneous magnetic field perpendicular to the substrate plane, as shown in Fig.\,\ref{fig:1}(a-c), generating a flux density $\mathbf{B}$. A transport current of density $j_\text{tr}$ is applied to the structures through a pair of normally conducting leads attached to the slit banks (strip edges), inducing an azimuthal current flow in both the open nanotube and the nanopetal. 

The cylindrical surfaces of the nanopetal and open nanotube are parameterized using 2D Cartesian coordinates, with the arc length in the azimuthal direction $x\in\left[\delta/2,2\pi R-\delta/2\right]$ and the coordinate along the cylinder axis $y\in\left[0,L\right]$. The flat nanostrip is described similarly; however, in this case, the spatial profile of $\mathbf{B}_n$ is homogeneous. All structures are assumed to be made from Nb films with a thickness of $d = 50$\,nm, allowing one to neglect finite-thickness effects\,\cite{Smirnova20}. Table \ref{table:parameters} summarizes the material parameters and their relations in the assumed dirty limit\,\cite{tinkham2004introduction}.

\begin{table}[b!]
    \centering
    \caption{Material parameters used in the TDGL simulations.}
    \begin{tabular}{p{3.5cm}p{2cm}p{2.5cm}}
    \hline
    \textbf{Parameters} & \textbf{Denotation} & \textbf{Value}\\
    \midrule
    Diffusion coefficient &$D$ &$1.2\times10^{-3}$\,m$^2$/s \\
    Normal conductivity &$\sigma$ &$16\;(\mu\Omega\mathrm{ m})^{-1}$ \\
    Relative temperature & $T/T_c$ & 0.95\\
    Penetration depth&$\lambda$ &273 nm \\
    Coherence length&$\xi$ &58 nm \\
    GL parameter&$\kappa=\lambda/\xi$ &4.7 \\
    \midrule
    \bottomrule
    \end{tabular}
    \label{table:parameters}
 \end{table}

The flat strip is modeled as a rectangle with dimensions $4.93 \times 5\,\mu$m$^2$ (width $W \times$ length $L$), while the open nanotube is modeled as a cylinder with radius $R = 0.8\,\mu$m and length $L = 5\,\mu$m, featuring a narrow slit with an arc width $\delta = 0.1\,\mu$m. The nanopetal is modeled as half of a cylinder with the same radius and length, with flat regions attached to both sides. The resulting width of the open nanotube and nanopetal is the same as that of the flat strip.

\subsection{Equations and parameters}
The spatiotemporal evolution of the superconducting order parameter $\psi$ in the nanostructures is described by the two-dimensional TDGL equation
\begin{equation}
\label{eGL}
    (\partial_t + i \varphi)\psi=\left(\nabla
    -i\textbf{A}\right)^{2}\psi
    +\left(1-|\psi|^{2}+U_{\text{\text{p}}}\right)\psi,
\end{equation}
where $\varphi$ is the scalar potential and $U_{\mathrm{p}}$ is the pinning potential\,\cite{key-34}. $\mathbf{A}$ is the vector potential tangent to the surface, which describes the normal component $B_{n} = \mathbf{B} \cdot \mathbf{n}$, with $\mathbf{n}$ being the unit vector normal to the (cylindrical) surface. In Eq.\,\eqref{eGL}, the differential operator $\nabla$ acts in the tangent space of the surface rather than in 3D space. The boundary conditions for the order parameter and the scalar potential read
\begin{subequations}
\label{eq:boundaries}
\begin{align}
    \psi=0,\quad
    \partial_x \varphi = - j_\mathrm{tr} / \sigma_\mathrm{n},
    \qquad
    \text{at contacts } \partial D_x,
    \\
    (\partial_y - i A_y) \psi = 0,\quad
    \partial_y \varphi = 0\quad
    \text{at free edges } \partial D_y.
\end{align}
\end{subequations}

\begin{table}[b!]
\caption{Dimensional units for quantities in Eqs.\,(\ref{eGL}) and (\ref{phij}) for Nb at $T/T_c = 0.95$.
\label{Table2}}
\begin{center}
    \begin{tabular}{p{3.5cm}p{2.5cm}p{2cm}}
 \hline
 \textbf{Parameter} & \textbf{Unit} & \textbf{Value} \\
 \midrule
 Time, $t$ & $\xi^2/D$ & 2.85 ps
 \\
 Length, $x$ & $\xi$ & 58 nm
 \\
 Magnetic field, $\mathbf{B}$ & $\Phi_0 / 2\pi\xi^2$ & 96 mT
 \\
 Current density, $\mathbf{j}$ & $\hslash c^2 / 8\pi\lambda^2 \xi e$ & 60 GA $\text{m}^{-2}$
 \\
 Electric potential, $\varphi$ & $\sqrt{2}H_c \xi \lambda / c \tau $ & 116 $\mu$V
 \\
 Conductivity, $\sigma$ & $c^2 / 4\pi\kappa^2D$ & 31 $(\mu\Omega \text{ m})^{-1}$
 \\
 \midrule
 \bottomrule
\end{tabular}\end{center}
\end{table}

The scalar potential satisfies the Poisson equation
\begin{equation}\label{phij}
    \nabla^{2}\varphi=\frac{1}{\sigma}\nabla\cdot\textbf{j}_{\text{sc}},\quad
\textbf{j}_{\text{sc}}=\mathfrak{Im}\left(\psi^{*}(\nabla - i\mathbf{A})\psi\right),
\end{equation}
where $\mathbf{j}_{\text{sc}}$ is the superconducting current density. The experimentally observable quantity is the voltage between the normally conducting leads, which is determined by the difference in scalar potentials at both leads, averaged over their length. Equation\,\eqref{eGL} is expressed in a dimensionless form using the units specified in Table\,\ref{Table2}. Link variables are employed to minimize errors arising from the system's gauge invariance. A finite-difference method with a grid resolution of $12.9 \times 12.9$\,nm and a time step of $0.025$\,ps \cite{key-25, key-26} is used to solve the TDGL equation. The data underlying the curves in the figures discussed in what follows are available at \cite{Bog25DSNonreciprocity}.

\subsection{Pinning sites}
\begin{figure}[t!]
    \centering
    \includegraphics[width=0.9\columnwidth]{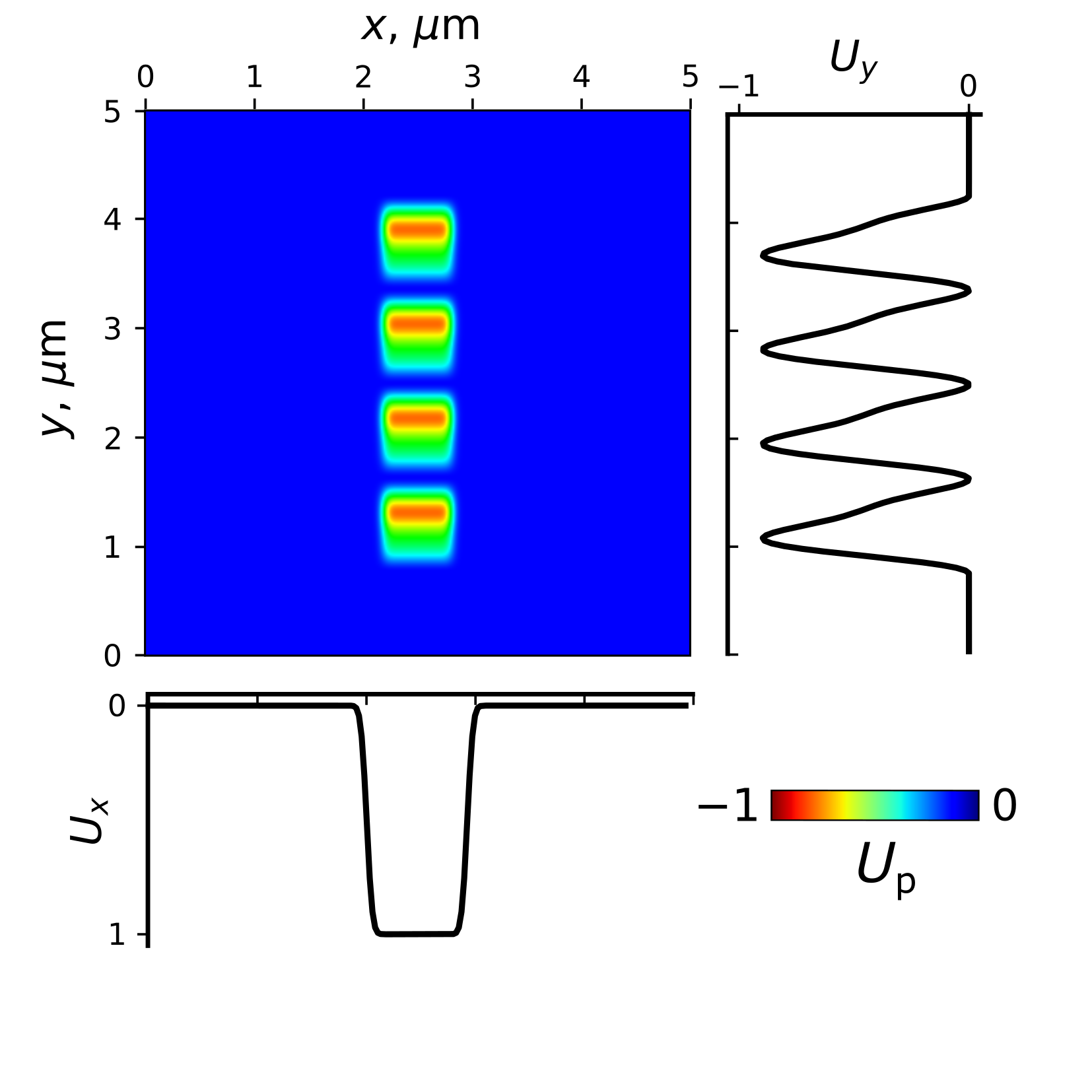}
    \caption{Coordinate dependences of the pinning potential $U_\text{p}(x,y)$ and its constituents $U_\text{x}(x)$ and $U_\text{y}(y)$.}
    \label{fig:2}
\end{figure}

We consider an array of asymmetric pinning sites, which induce a dimensionless pinning potential $U_{\text{p}}$ described by the following expression, as can be realized experimentally, e.g., for Nb films with asymmetric nanogrooves\,\cite{Dob17scr},
\begin{align}
\label{ePP}
    &
    U_{\text{p}}= U_\text{x}(x) U_\text{y} (y),
    \\\nonumber&
    U_\text{x}(x) =
    \frac{1}{4}
    \left(
        1 + \text{erf}\;\frac{2x - W + W_\text{U}}{2\xi}
    \right) \left(
        1 + \text{erf}\;\frac{W + W_\text{U} - 2x}{2\xi}
    \right),
    \\\nonumber&
    U_\text{y}(y) = \frac{U_0}{2} \left(
        1 - \cos(\varphi(y)) + 0.28 (1 - \sin(2\varphi(y)))
    \right) - a,
    \\\nonumber&
    \varphi(y) = ky + b,
\end{align}
Here, $W_\text{U}$ is the width and $U_0$ is the depth of the pinning potential well, while the constants $b$ and $k$ define the geometric shift and period of the pinning array along the $y$-coordinate. The constant $a$ ensures that the potential is zero between the pinning sites. The pinning site width is chosen as $1/8$ of the full membrane width, $W_\text{U} = W / 8 \approx 0.6\,\mu$m, and $U_0 = -0.8$, resulting in a potential minimum of approximately $-0.9$. For all considered nanoarchitectures, four pinning sites are placed at the center of the membrane, as shown in Fig.\,\ref{fig:2}. Although the number of pinning sites is chosen arbitrarily, it is essential for our study that this number is the same across all structures. Additionally, the pinning sites are positioned sufficiently far from the strip edges, minimizing any potential edge-barrier suppression effects, and the normally conducting leads are placed at an equal distance from the pinning site array. 

\section{Results}

\begin{figure}[t!]
    \includegraphics[width=0.3\textwidth]{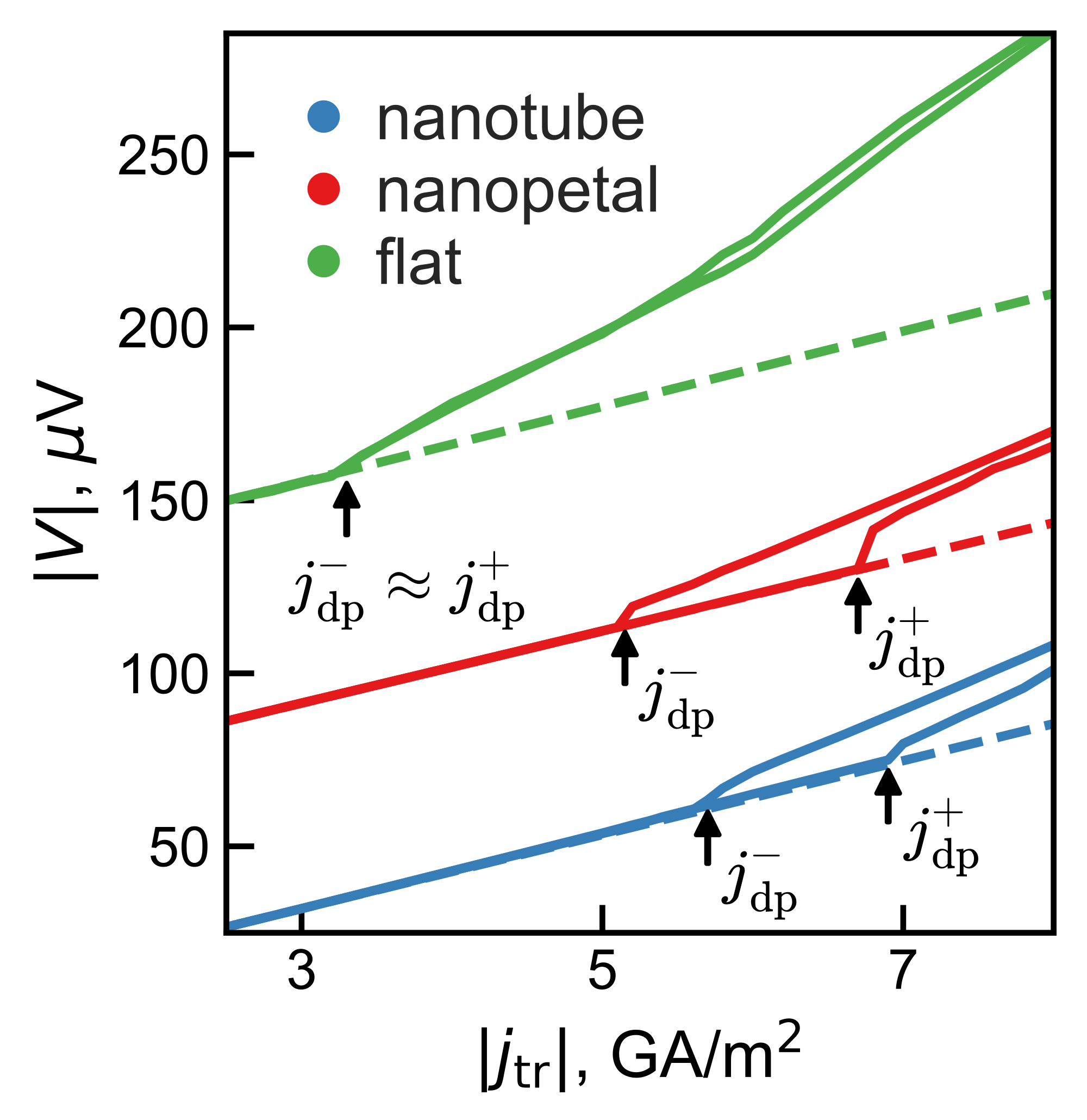}
    \caption{CVCs for both current directions in the nanostructures at $6$\,mT. The arrows indicate the depinning current density. A larger depinning current corresponds to vortex motion in the harder direction of the asymmetric pinning potential. The CVCs are shifted upward with respect to each other by 60\,$\mu$V for better readability.}
    \label{fig:3}
\end{figure}

Figure\,\ref{fig:3} shows the CVCs of all nanostructures as the transport current density $j_\mathrm{tr}$ increases for both polarities. Note that, in the considered structures, the initial linear increase in voltage corresponds to immobile, pinned vortices. This voltage arises due to the resistance of the areas near the leads, resulting from the proximity effect. A notable increase in the CVC slopes with increasing $j_\mathrm{tr}$ indicates the depinning of vortices. Vortex depinning depends not only on the transport current density, magnetic field, and pinning strength, but also on the presence or absence of nearby vortices. Therefore, vortex depinning is a probabilistic event, and not all vortices in the nanostructure depin at the same value of $j_\text{tr}$\,\cite{Sil12njp}. In our analysis, the quantities $j^\pm_\text{dp}$ represent the transport current density values at which at least one vortex is depinned. When all vortices are depinned, the resistance increases linearly, indicating a regime of free flux flow. In summary, three regions in the CVCs are identified: pinned vortices, moving vortices, and the transition between these two regimes.

\begin{figure}[t!]
    \includegraphics[width=0.6\linewidth]{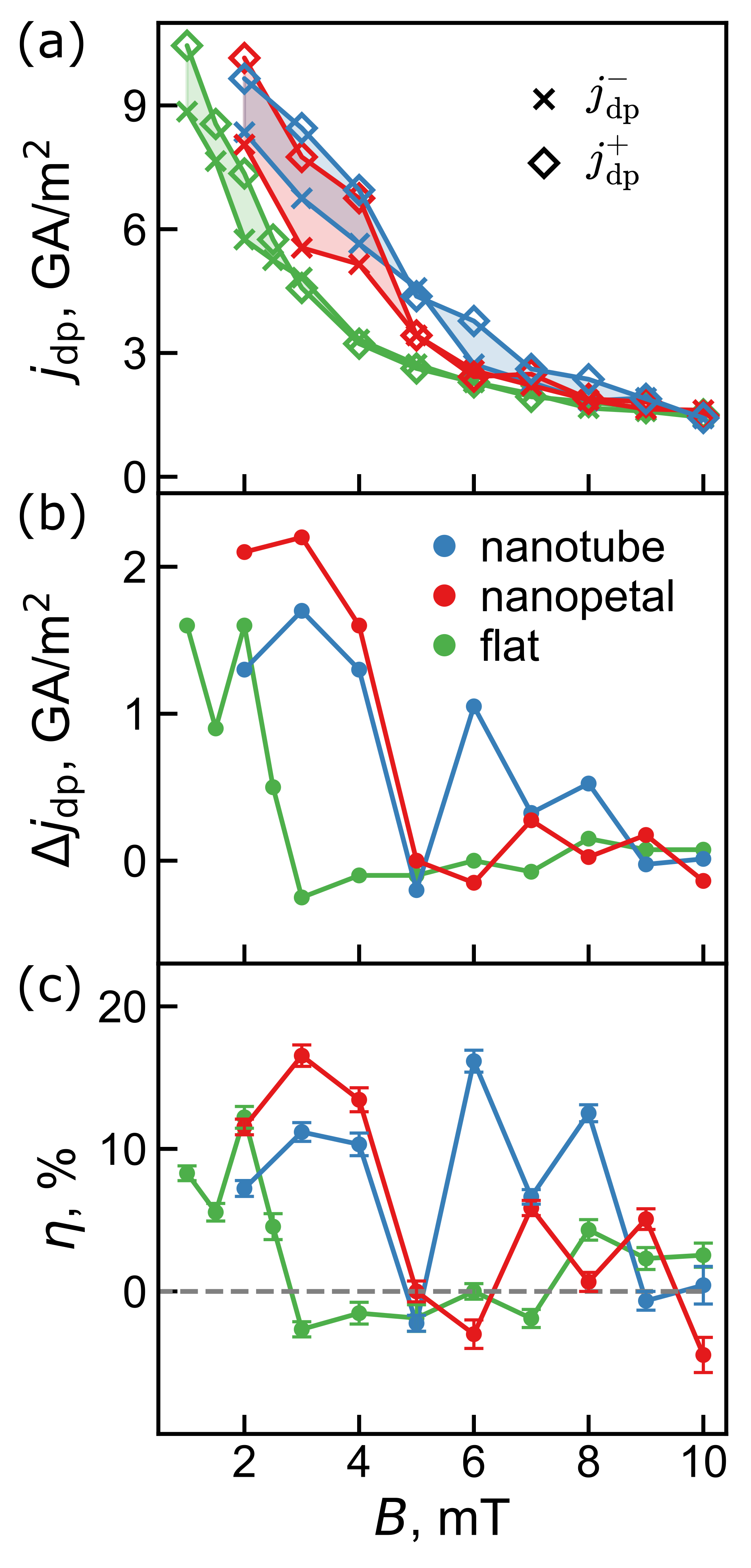}
    \caption{(a) Depinning current density $j_\text{dp}$, (b) the difference in depinning current density $\Delta j_\text{dp}$ between the hard and easy directions, and (c) the ratchet effect efficiency $\eta$ as functions of the magnetic field $B$ for the planar structure, nanopetal, and open nanotube.}
    \label{fig:4}
\end{figure}

To quantify the VRE efficiency, we compare
\begin{equation}
    \eta \equiv\frac{j^+_\text{dp}-j^-_\text{dp}}{j^+_\text{dp}
    +j^-_\text{dp}}
\end{equation} 
as a function of $B$ for all nanostructures at $T/T_c = 0.95$. The results are presented in Fig.\,\ref{fig:4}. The difference between $j^\pm_\text{dp}$ is significant for the planar structure at low magnetic fields ($\lesssim 2.5$\,mT), but it disappears at higher fields. The nanopetal and nanotube exhibit a difference in $j^\pm_\text{dp}$ up to about $4$\,mT and $7$\,mT, respectively. Additionally, in the open nanotube, the depinning current difference vanishes at approximately $5$\,mT.

\begin{figure}[t!]
\includegraphics[width=0.99\linewidth]{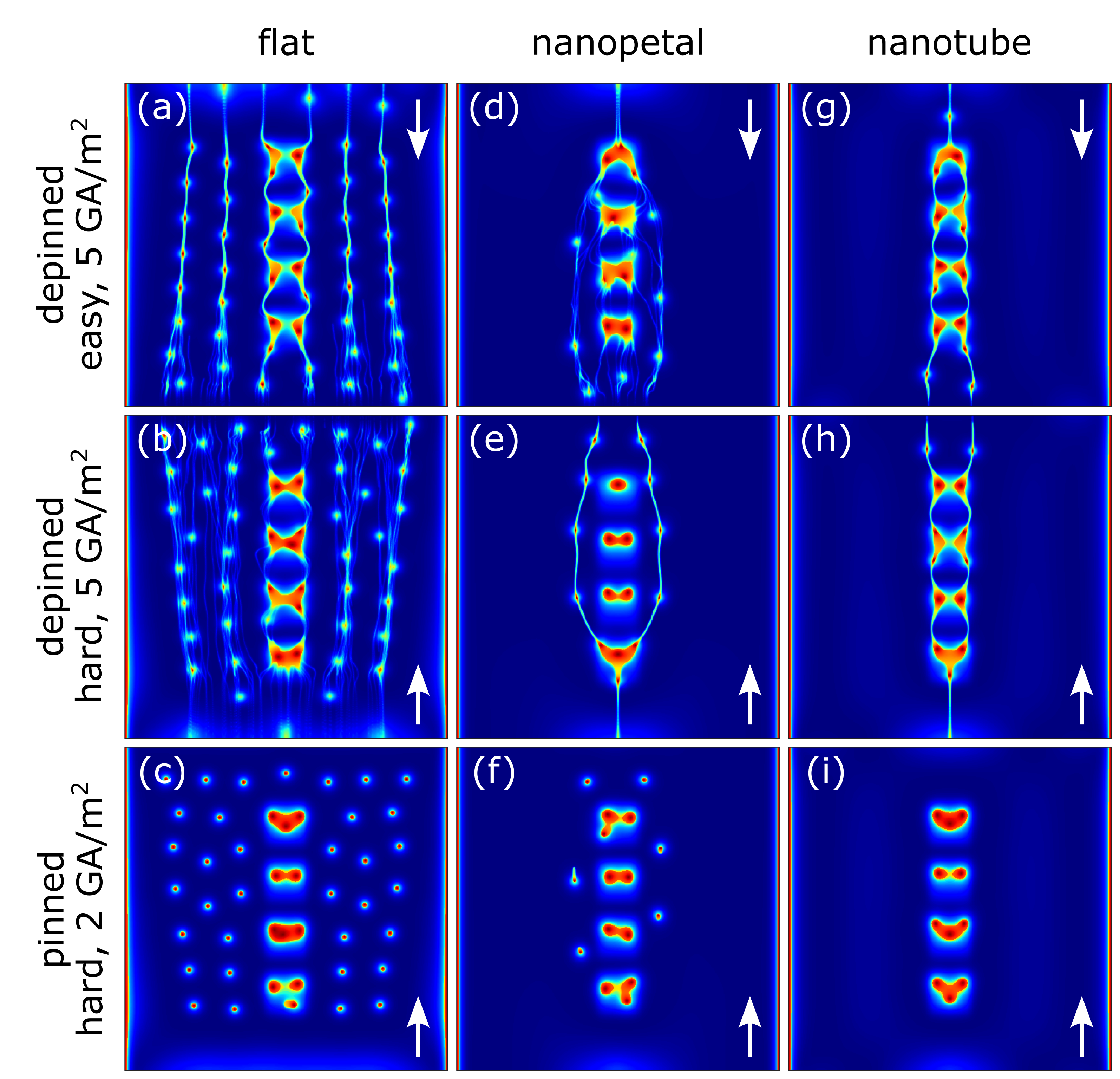}
\caption{Vortex trajectories superimposed on snapshots of the spatial distribution of the modulus of the superconducting order parameter $|\psi|$ in the flat membrane (a-c), nanopetal (d-f), and nanotube (g-i) at $6$\,mT. The first row shows depinned vortices moving in the easy direction at $5$\,GA/m$^2$; the second row shows depinned vortices moving in the hard direction at $5$\,GA/m$^2$; and the third row shows pinned vortices at a transport current density of $2$\,GA/m$^2$, pushing the vortices to move in the hard direction. Arrows indicate the direction of the driving Lorentz force.}
\label{fig:5}
\end{figure}

\section{Discussion}
\paragraph*{Enhancement of the ratchet effect in 3D nanostructures.}
The key result of our study is that the VRE persists at higher magnetic fields for the open nanotube and nanopetal compared to the planar strip, even though the number of pinning sites is the same across all structures. The vortex trajectories in the structures, shown in Fig.\,\ref{fig:5}, provide the following explanation for the enhanced VRE. In the nanotube and nanopetal, vortices move through regions where $B_n$ is maximal, and the asymmetric pinning sites along the vortex trajectories further enhance the persistence of the VRE. A similar channeling effect was discussed in our previous work\,\cite{Bog24prb} in the context of non-diverging vortex jets in 3D open nanotubes. For the planar strip, $B_n$ is homogeneous, and there are no preferred areas for vortex motion that could be influenced by magnetic field inhomogeneity. Of course, the vortex trajectories are influenced by both vortex-vortex and vortex-pinning interactions. However, for the planar strip, additional rows of pinning sites would be required to impede the vortex motion across the entire structure.

The VRE in the nanopetal vanishes for $B \gtrsim 6$\,mT (see Fig.\,\ref{fig:4}(d-f) and Fig.\,\ref{fig:4}(g-i)). An inspection of the vortex trajectories in Fig.\,\ref{fig:5} reveals that, despite the identical geometry in the middle of the nanopetal and nanotube, at $6$\,mT, vortex focusing in the arc of the nanopetal is much weaker than in the upper part of the open nanotube. We attribute this difference to the distinct configurations of the screening currents in both structures. Specifically, in the nanopetal, the screening current forms a single contour, while in the nanotube, it consists of three contours with alternating circulation directions. In the latter case, the screening current is squeezed into a narrower region at the top of the nanotube, thereby enhancing the vortex focusing effect.

To enhance the VRE, the nanotube radius must be carefully chosen, considering two key effects. The first is the focusing effect, which becomes stronger as the tube radius decreases. The second effect is the increase in the transport current density required for vortex nucleation as the tube radius decreases. For nanotubes with a radius of $390$\,nm, vortex nucleation occurs at $\gtrsim 8$\,GA/m$^2$ for $5$\,mT\,\cite{bog25shapiro}, which exceeds the depinning current in the present study. Consequently, no non-reciprocity is observed, as there are no vortices available for depinning at the corresponding transport current densities. Thus, the predicted high VRE efficiency in nanotubes with a realistic radius of $800$\,nm makes them promising candidates for experimental investigations of the VRE.

\paragraph*{Applicability of the model.}
Although the nanoarchitectures under consideration are 3D objects, their numerical modeling here relies on an effectively 2D TDGL model. The applicability of the 2D model is justified as long as the membranes are thin enough to neglect finite-thickness effects. First, the film thickness $d$ must be smaller than the penetration depth, $d < \lambda$. This condition ensures that the induced magnetization does not significantly affect the order parameter dynamics. Second, the magnetic field component tangent to the surface can lead to vortex nucleation with cores along the cylinder walls in the paraxial direction, which is suppressed when the thickness is smaller than a few $\xi(T)$, with $2\xi(T)$ providing an estimate for the vortex core diameter. Third, large surface curvatures lead to curvature-induced potentials of the da Costa type, which are negligible if the curvature radius $R$ is much larger than the thickness $d$ and coherence length $\xi$\,\cite{da1981quantum,jensen1971quantum}.

\paragraph*{Possible experimental examination.}
Superconductor 3D nanoarchitectures can be fabricated by strain-relaxation-driven self-rolling\,\cite{Thurmer08,Loe19acs}, with artificial pinning sites created using focused ion and electron beams\,\cite{Fer20mat,Vol24nac}. The vortex dynamics in an asymmetric pinning potential require weak bulk pinning due to the intrinsic disorder in the superconducting membrane. Accordingly, the predicted effects should be experimentally examined in superconductors with weak background pinning, while the asymmetric pinning sites should be strong pins. For instance, the direct-write superconductor Nb-C satisfies these conditions\,\cite{Dob20nac}, and 3D nanostructures made of this material have already been demonstrated\,\cite{Por19acs}. Additionally, strong pinning sites can be realized through focused ion beam milling\,\cite{Dob15met} or by ferromagnetic decoration assisted by focused ion or electron beams\,\cite{Dob10sst}.

\section{Conclusion}
We have analyzed the vortex ratchet effect in 3D open nanotubes, nanopetals, and planar strips with an array of asymmetric pinning sites. Our analysis, based on the numerical solution of the TDGL equation, reveals that the vortex ratchet effect persists at higher magnetic fields in the 3D nanostructures compared to the 2D planar strip. The enhanced vortex ratchet effect in the 3D nanostructures is attributed to the combined effects of the inhomogeneous magnetic field and the asymmetry of the pinning sites. Specifically, the vortices move through regions where the normal component of the magnetic field is maximal, forcing them to channel along areas with asymmetric pinning sites.

\section*{Acknowledgements}
    The work of I.B. was funded by the Deutsche Forschungsgemeinschaft (DFG, German Research Foundation) under Germany’s Excellence Strategy –- EXC-2123 QuantumFrontiers –- 390837967. I.B. gratefully acknowledges the use of the CryoMind simulation workstation at CryoQuant/TU Braunschweig. R.H.dB. thanks the research funding agency CAPES for the Sandwich Ph.D. scholarship, the IFW for hospitality, and J. Albino Aguiar for his support. V.M.F. gratefully acknowledges the NHR Center NHR@TUD for the provided computing time and M.D. Croitoru for useful discussions. The research is supported by the European Cooperation in Science and Technology COST Action CA21144 (SuperQuMap).

\section*{Conflict of Interest}
The authors declare no conflict of interest.

\section*{Keywords}
Superconductivity, 3D nanoarchitectures, vortex dynamics, pinning, vortex ratchet effect.

\section*{Data availability statement}
The data that support the findings of this study are openly available in Mendeley Data\,\cite{Bog25DSNonreciprocity}.

\bibliography{main}

\end{document}